\documentclass{aa}  

\usepackage{graphicx}
\usepackage{hyperref}
\usepackage{txfonts}
\usepackage{natbib}
\usepackage{ulem}
\usepackage{xcolor}
\usepackage{diagbox}
\usepackage{multirow}

\definecolor{seagreen}{rgb}{0.190, 0.525, 0.361}
\definecolor{cerulean}{rgb}{0.165, 0.322, 0.745}
\definecolor{goldenrod}{rgb}{0.855, 0.647, 0.125}

\newcommand{\REV}[1]{{#1}}

\newcommand{\msun}{{{\rm M}_\odot}}

\begin{document} 
   \title{Isles of regularity in a sea of chaos amid the gravitational three-body problem}


   \author{Alessandro Alberto Trani
	\inst{1,2,3}
	\and
	Nathan W.C. Leigh\inst{4,5}
	\and
	Tjarda C.N. Boekholt\inst{6}
	\and
	Simon Portegies Zwart\inst{7}}

	\institute{Niels Bohr International Academy, Niels Bohr Institute, Blegdamsvej 17, 2100 Copenhagen, Denmark
	\and Research Center for the Early Universe, School of Science, The University of Tokyo, Tokyo 113-0033, Japan
	\and Okinawa Institute of Science and Technology, 1919-1 Tancha, Onna-son, Okinawa 904-0495, Japan   
	\email{aatrani@gmail.com}
	\and
	Departamento de Astronom\'ia, Facultad Ciencias F\'isicas y Matem\'aticas, Universidad de Concepci\'on, Avenida Esteban Iturra, Casilla 160-C, Concepci\'on, 4030000, Chile
	\and
	Department of Astrophysics, American Museum of Natural History, Central Park West and 79th Street, New York, 10024, NY, USA
	\and
	NASA Ames Research Center, Moffett Field, 94035, CA, USA
	\and
	Leiden Observatory, Leiden University, PO Box 9513, 2300 RA, The Netherlands
	\\}

   \date{Received M DD, YYYY; accepted M DD, YYYY}

  \abstract
{The three-body problem (3BP) poses a longstanding challenge in physics and celestial mechanics. Despite the impossibility of obtaining general analytical solutions, statistical theories have been developed based on the ergodic principle. This assumption is justified by chaos, which is expected to fully mix the accessible phase space of the 3BP.}
{This study probes the presence of regular (i.e. non-chaotic) trajectories within the 3BP and assesses their impact on statistical escape theories.}
{Using three-body simulations performed with the accurate, regularized code \textsc{tsunami}, we established criteria for identifying regular trajectories and analysed their impact on statistical outcomes.}
{Our analysis reveals that regular trajectories occupy a significant fraction of the phase space, ranging from 28\% to 84\% depending on the initial setup, and their outcomes defy the predictions of statistical escape theories. The coexistence of regular and chaotic regions at all scales is characterized by a multi-fractal behaviour. Integration errors manifest as numerical chaos, artificially enhancing the mixing of the phase space and affecting the reliability of individual simulations, yet preserving the statistical correctness of an ensemble of realizations.}
{Our findings underscore the challenges in applying statistical escape theories to astrophysical problems, as they may bias results by excluding the outcome of regular trajectories. This is particularly important in the context of formation scenarios of gravitational wave mergers, where biased estimates of binary eccentricity can significantly impact estimates of coalescence efficiency and detectable eccentricity.}


   \maketitle
%

\section{Introduction}
The gravitational three-body problem (3BP) serves as an exemplary case of chaos in nature, showcasing the complexity that can emerge in seemingly simple systems \citep{poincare_newmeth}. The inherently chaotic nature of the 3BP makes it challenging to find analytic solutions, with the only existing general solution being impractical because of slow convergence \citep{sundman1912,belorizky1930,barrowgreen2010}, and all other solutions being limited to periodic orbits  \citep{broucke1975a,broucke1975b,hadjidemetriou1975a,hadjidemetriou1975b,henon1976,moore1993,chenciner2000,suvakov2013} or to the hierarchical configuration \citep{kinoshita1999,kinoshita2007}.

Even obtaining individual solutions for specific realizations of the non-hierarchical 3BP is a far more complex task than it might initially appear. We define the non-hierarchical 3BP as the scenario in which three bodies of comparable masses (i.e. $m_1 \sim m_2 \sim m_3$, as opposed to the restricted 3BP where $m_1 \ll m_2, m_3$) engage in interactions within a confined region of space, exchanging energy and angular momentum (here we exclude braids and other periodic solutions). In the point-particle limit, these complex interactions continue until one particle is ejected, leaving behind a bound binary system \citep{agekyan1971,szebehely1971,standish1972a}. 

\begin{figure*}
	\sidecaption
	\includegraphics[width=\linewidth]{{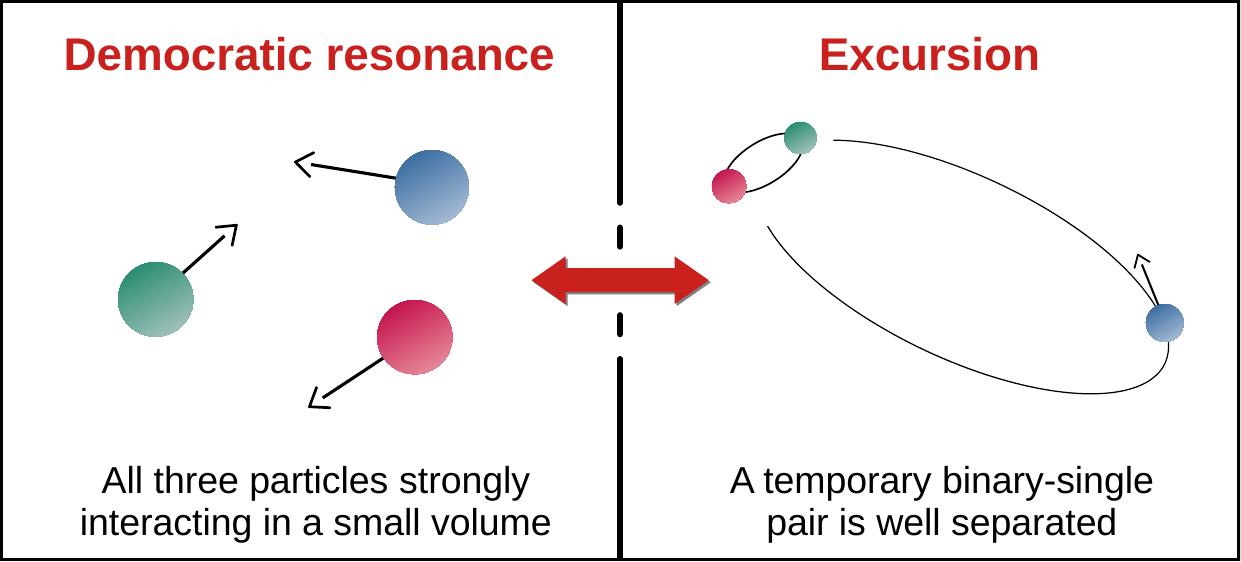}}
	\caption{Two states of evolution in a non-hierarchical three-body interaction. The system cycles between the two states, until it breaks up into an unbound binary-single pair.
	}
	\label{fig:schema}
\end{figure*}

No matter how complex a three-body interaction may appear, its evolution can be subdivided into two distinct states between which the system alternates (see Figure~\ref{fig:schema}). The first state occurs when all three particles interact within a confined region of space, devoid of any fixed hierarchy among them. We refer to this state of the interaction as a `democratic resonance', as the particles freely exchange energy and angular momentum among each other \citep[see][]{hegg93}. Democratic resonances cannot persist indefinitely, because this configuration is inherently unstable, and the system eventually forms a temporary hierarchy, transitioning into the second state, which we term `excursion'. Excursions occur when two bodies form a temporary binary and the third body recoils while remaining bound to the binary centre of mass. In this state, the system forms an unstable, temporary hierarchical triple, which reverts back to the democratic resonance state once the single body returns and interacts with the binary. 
A three-body interaction cycles between these two states until the system finally breaks up into an unbound binary-single pair. The breakup can only occur from the democratic resonance state, because during an excursion, the system can be considered as non-interacting. 
The physical origin of chaos in the non-hierarchical 3BP arises from the close gravitational interactions during the democratic resonances.

To map initial conditions (positions and velocities) to the final state of the system (ejection velocity vector of the single particle, energy, and angular momentum vector of the binary), one must numerically integrate the equations of motion until the three-body system decays into a binary and a single. In the best-case scenario, such a numerically approximated solution is sufficiently accurate and faithful to the underlying intrinsic solution. However, even with the most sophisticated computational techniques, limitations arise due to the accuracy of the integration algorithm, the constraints of floating-point arithmetic, and available computational resources. Consequently, even small integration errors can propagate and magnify over time, inevitably leading to inaccurate results.

Fortunately, it is possible to leverage chaos to find statistical solutions for the 3BP, also called statistical escape theories, which have an origin in the pioneering work of \citet{mon76a}. His approach, which is based on the microcanonical ensemble of statistical mechanics, laid the foundation to explore the interplay between deterministic chaos and probabilistic dynamics in celestial mechanics.

The key idea of statistical escape theories is that chaotic interactions mix the phase space of the 3BP, so that over the lifetime of the system, every available state is eventually visited, filling the phase space in a uniform and random sense. Except for conserved quantities (the energy and angular momentum vector), the system has lost memory of its initial conditions due to chaos and, at any given time, it can be found at any point in the accessible phase space. This is the definition of an ergodic system in statistical mechanics, meaning that the time average of some property over the trajectory is equal to the ensemble average over the phase-space volume \citep{birkhoff1931}. In the case of the non-hierarchical 3BP, this means that the probability of the outcome properties are proportional to the phase space that is accessible at the breakup time of the triple system. Given the energy and angular momentum vector, it is then possible to characterize and sample the accessible phase space at the moment of breakup to obtain the distributions of the final outcome properties \citep{mon76a,mon76b}.

Statistical escape theories typically provide the closed-form expression for the tri-variate distribution $f(E_{\rm B}, L_{\rm B}, i_{\rm S})$ of the outcome properties, expressed in terms of final binary binding energy $E_{\rm B}$, angular momentum $L_{\rm B}$, and the angle $i_{\rm S} = \arccos(\vec{L}_{\rm S} \cdot \vec{L}_{\rm 0})$ between the total angular momentum $\vec{L}_0$ and the angular momentum of the ejected single $\vec{L}_{\rm S}$. The original statistical escape theory by \citet{mon76a} has been improved and refined over the years, yielding a remarkably good agreement with the results of numerical experiments \citep{stan72,sasl74,valt74,mon76b,valt05,stone2019,manwadkar2020,manwadkar2021,ginat2021}.

However, statistical escape theories suffer from the assumption that the phase space of the 3BP is fully mixed. It has been highlighted that the 3BP is not fully chaotic, and its phase space is divided between chaotic and regular (i.e. non-chaotic) trajectories \citep{shevchenko1998a,shevchenko1998b,shevchenko2010,manwadkar2020,parischewsky2023}. \REV{Therefore, we can expect that the predictions of the statistical escape theories clash with the outcomes of regular 3BP interactions. Only recently, a new statistical theory of the 3BP, not based on the microcanonical ensemble, has been proposed with the aim of solving this issue \citep{kol2021}. Simulations show an improved agreement of the predicted ejection probabilities with the numerical experiments, compared to previous theories \citep{manwadkar2021,manwadkar2023}.
}

In this paper, we explore the presence of regular trajectories in the 3BP and its impact on the statistical predictions of the 3BP escape theories. Section~\ref{sec:meth} details the setup of our simulations and our numerical model. Criteria for identifying regular trajectories in the initial phase space are outlined in Section~\ref{sec:stats}, which also illustrates the influence of regular regions on the statistical outcomes of the 3BP. Section~\ref{sec:fractals} delves into the fine-grained mixing between regular and chaotic regions occurring at all scales, characterizing the multi-fractal nature of this mixing. Integration errors and their role in spurious phase-space mixing, referred to as numerical chaos, are examined in Section~\ref{sec:numchaos}. Section~\ref{sec:planck} highlights the extreme sensitivity of the 3BP to perturbations below the Planck length, where mixing between regularity and chaos persists. In Section~\ref{sec:astro} we highlight the astrophysical implications of our work. Finally, our conclusion and outlooks are summarized in Section~\ref{sec:conclus}.

\section{Numerical methods}\label{sec:meth}
\begin{figure*}
	\sidecaption
	\includegraphics[width=\linewidth]{{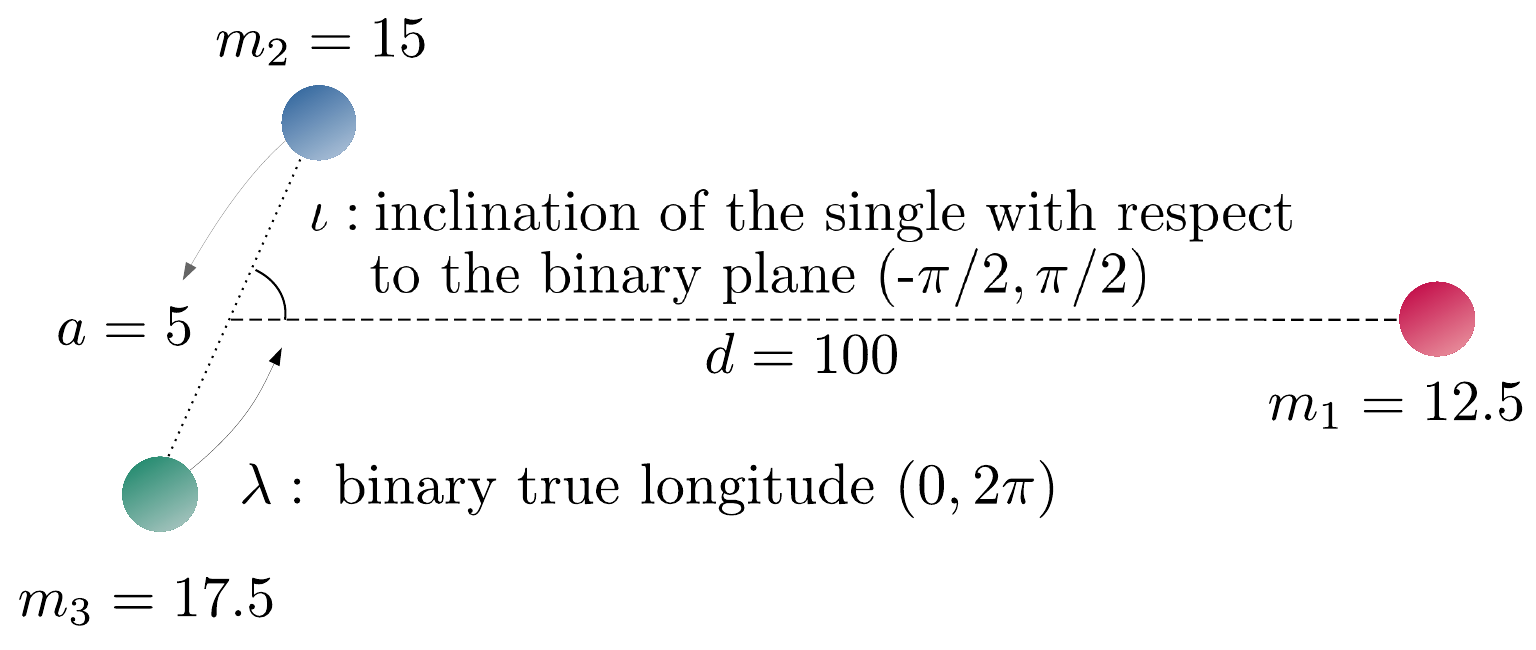}}
	\caption{Initial setup of our suite of simulations. We restrict ourselves to two degrees of freedom: the initial inclination $\iota$, and the binary phase (true longitude) $\lambda$. The reference units are au and solar masses. We simulate $10^6$ realizations for a total integration time of $10^9 \rm\, yr$.
	}
	\label{fig:setting}
\end{figure*}

We perform a suite of numerical simulations of three-body interactions. The interactions begin with a binary with a fixed semi-major axis of $a = 5\rm\,au$ and a single, which undergo an interaction with a given total energy, angular momentum and particle masses. The initial setup is illustrated in Figure~\ref{fig:setting}. The centre of mass of the binary and the single are at rest in the centre-of-mass frame of the system, at a distance of $d = 100 \,\rm au$.

We restrict the degrees of freedom of the initial conditions to the 2-dimensional space consisting of the binary true longitude $\lambda$ and its inclination $\iota$ with respect to the line joining it with the single. The initial binary is circular.

We sample the initial parameter space uniformly in the $\lambda$-$\iota$ space, as shown in Figure~\ref{fig:bigfig}. \REV{We simulate $10^6$ realizations for the whole parameter space ($\iota \in [-\pi/2, \pi/2]$, $\lambda \in [0,2\pi)$). We also perform a series of 14 zoom-ins into the $\lambda$-$\iota$ space, by sampling $10^5$ initial conditions in 1/25th of the area of the previous zoom level (i.e. a shrinking factor of 5 in each dimension). Both $\lambda$ and $\iota$ are sampled uniformly. However, when calculating the statistical properties of the sets, such as the fraction of regular interactions, we weight each realization by $\cos{\iota}$ to ensure that our statistics are consistent with $\iota$ being uniform in $\sin{\iota}$}.

We integrate the motion of the three particles assuming no particle size and no collision, by means of the \textsc{tsunami} integrator \citep{trani2019a,trani2023iaus}. \textsc{tsunami} uses a combination of algorithmic techniques \citep{stoer1980,mik93,mik99a} to improve the accuracy and speed of the $N$-body integration, which results in total energy errors $<10^{-10}$.

We also keep track of the state of the simulation using the same classification scheme outlined in \citet{manwadkar2023} and \citet{trani2023}. Namely, we stop the simulation when a single body is on a hyperbolic, diverging orbit with respect to the remaining binary, and we record this time as the lifetime $t_{\rm life}$ (alternatively, the disruption time) of the triple system. The simulations are integrated until either the system breaks up, or until we reach a total integration time of $10^9\rm\,yr$. Only 6 simulations out of $10^6$ have not reached the triple breakup by the final integration time.

\section{Regular trajectories elude statistical escape theories}\label{sec:stats}

\begin{figure*}
	\centering
	\includegraphics[width=\textwidth]{{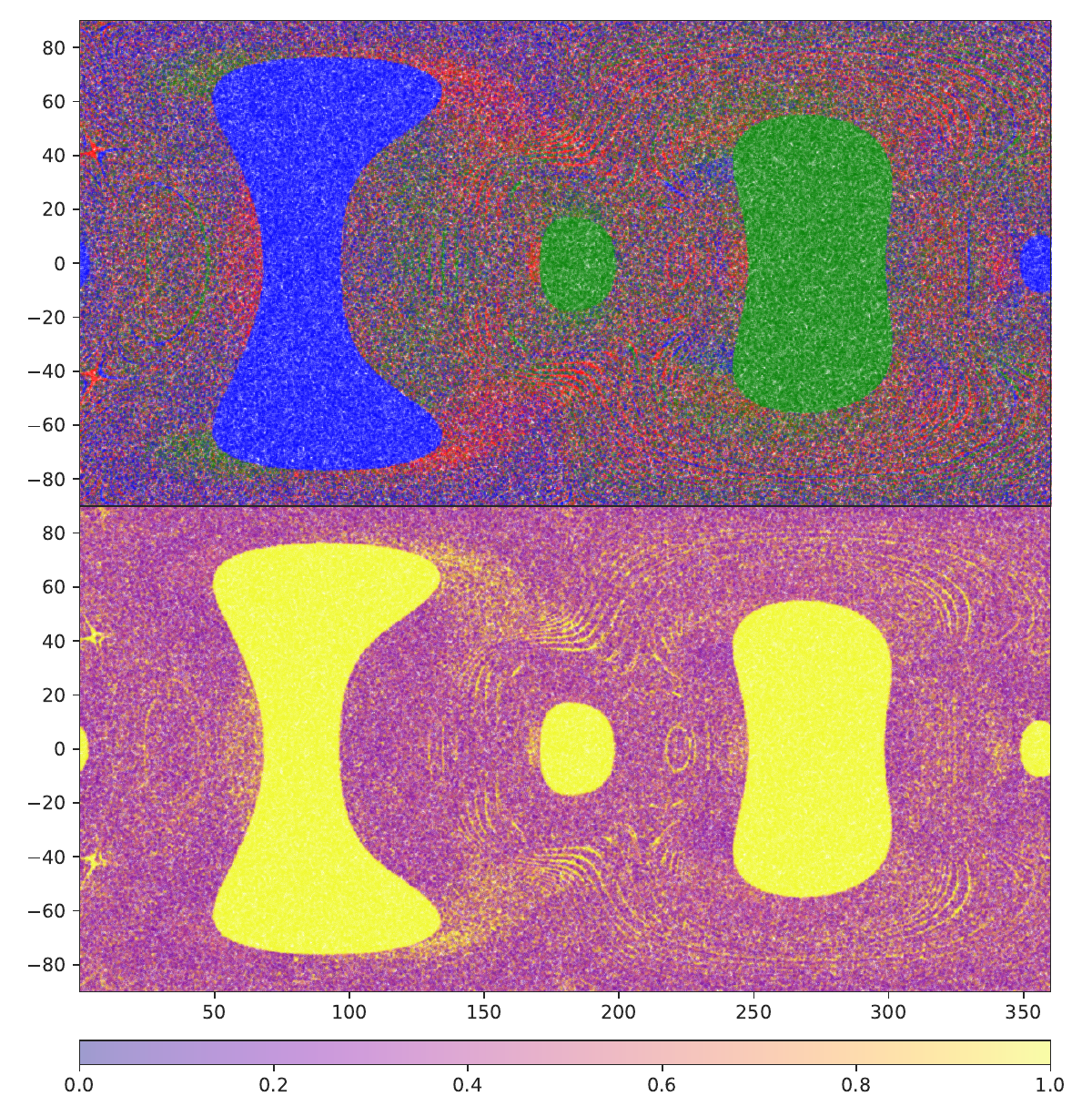}}
	\caption{Initial condition space in inclination $\iota$ ($y$-axis) and true longitude $\lambda$ ($x$-axis). Each dot represents an individual realization among $10^6$. Top panel: colour-coded by the identity of the escaping particles. Red: $12.5 \,\rm M_\odot$. Blue: $15 \,\rm M_\odot$. Green: $17.5 \,\rm M_\odot$. Regions of sparse density may appear as white dots due to the white background. Bottom panel: fraction of same-colour neighbouring particles $f_{\rm col}$, out of $k=12$ nearest neighbours.
	}
	\label{fig:bigfig}
\end{figure*}

The top panel of Figure~\ref{fig:bigfig} shows the initial phase space map, colour-coded by the identity of the escaping particle. In this space, neighbouring points share similar initial conditions. If the 3BP was fully ergodic, we should see a mix of colours everywhere, but instead we observe four large regions of uniform colours, two large ones at $\iota \sim 80^\circ$ and $260^\circ$, and two small ones at $\iota \sim 175^\circ$ and $\iota \sim 355^\circ$, that we term regular islands. These regular islands are constituted by regular trajectories (as opposed to chaotic trajectories), where the final state of the interaction smoothly maps onto the initial conditions.
These correspond to escapes of either of the two members of the binary. The blue and green regular islands present a translational quasi-symmetry by 180$^\circ$ along the $\lambda$-axis, which correspond to the binary phase swapping the two binary members. The slightly unequal masses prevent it from being a perfect symmetry, and make the small green island (escape of the most massive $m_1 = 17.5\,\msun$ particle) larger than the small blue island, and the large blue island more lopsided at high $\iota$ than the green one.

\begin{figure*}
	\centering
	\includegraphics[width=1.0\textwidth]{{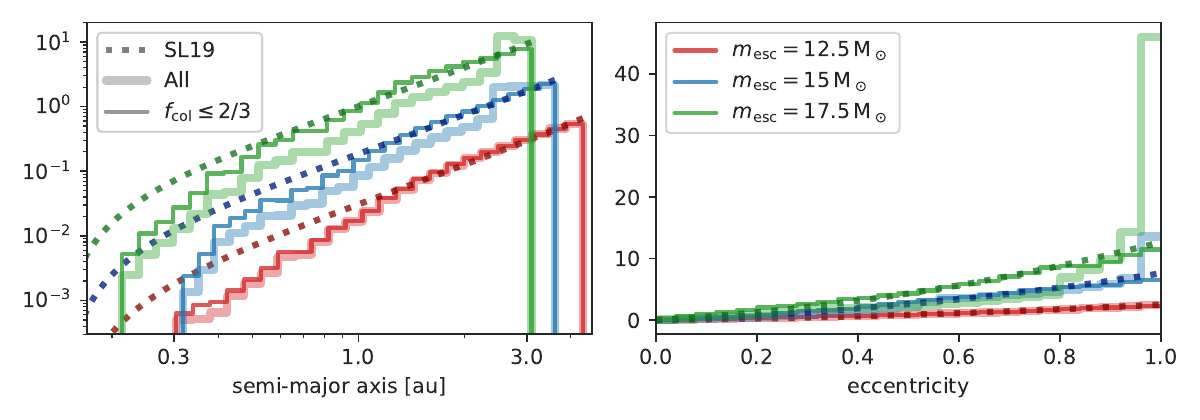}}
	\caption{Distributions of semi-major axes (left panel) and eccentricity (right panel) of the final binary for the entire initial phase space (i.e. the one in Figure~\ref{fig:bigfig}). The colours indicate the identity of the escaped particle, while the dashed lines indicate the predictions of the statistical escape theory of \citet{stone2019}. Thin lines: considering all simulations. Thick lines: excluding the regular islands ($f_{\rm col}\leq2/3$). In the left panel, we magnified the differences in semi-major axes distributions among different escaper by introducing different normalizations ($\times0.9$ for red, $\times3$ for blue, $\times20$ for green). In the right panel, the analytic predictions for the eccentricity distributions for different escapers would be virtually identical, so we introduced different normalizations ($\times3$ for blue, $\times5$ for green) to better distinguish the coloured lines.
	}
	\label{fig:comparedist}
\end{figure*}

To quantify how much area is occupied by the regular regions, we employ a $k$ nearest neighbour classification scheme.
For each point in the phase space, we calculate the fraction $f_\mathrm{col}(k)$ of $k$ neighbours that have the same colour as the point in consideration. The bottom panel of Figure~\ref{fig:bigfig} colour-codes the realizations based on $f_{\rm col}$ for $k=12$.  This criterion appears to effectively identify the major regular islands while also revealing numerous narrow, fine-grained regular regions outside these islands, which we refer to as regular stripes. 

If we define the regular trajectories as those that have $f_{\rm col}>2/3$, we find that about 37\% of the phase space is constituted by regular regions, with the blue ($15 \,\rm M_\odot$) and green ($17.5 \,\rm M_\odot$) regular regions constituting about 14\% and 13\% of the phase space each, and the remaining 9\% constituted by escapes of the lighter body (red, $12.5 \,\rm M_\odot$). The proportion of regular trajectories increases with the mass of the particles. This is intuitively expected, as for $m_3/m_1, m_3/m_2 \ll 1$, the motion of the binary becomes a Keplerian orbit (and thus regular by definition), while the motion of the point mass remains chaotic.

As demonstrated by \citet{manwadkar2021}, regular trajectories exhibit ejection probabilities that deviate from the predictions of statistical escape theories. For instance, if we consider all the simulations of Figure~\ref{fig:bigfig}, the ejection probabilities for bodies of mass $12.5$, $15$, and $17.5\,\rm M_\odot$ are found to be 43\%, 30\%, and 27\%, respectively. In contrast, according to the statistical escape theory of \citet{stone2019}, these probabilities are expected to be 64\%, 25\%, and 11\%, given our initial energy and angular momentum.

An unexplored aspect thus far is how these regular trajectories influence the distribution of outcome properties. If we were to colour-code the outcome properties such as the final binary semi-major axis in the initial phase space map, smooth transitions would appear corresponding to the regular regions. It is therefore not surprising that such regular regions have a considerable impact on the outcome properties, as shown in Figure~\ref{fig:comparedist}, which compares the marginalized distributions of semi-major axis and eccentricity to the theoretical predictions of \citet{stone2019}. 

The distributions of final outcomes, when considering all realizations, are at odds with the theoretical expectations for the ejected masses of $m_{\rm ej} = 15$ and $17.5 ,\rm M_\odot$ (blue and green lines). The discrepancy with the theory is particularly visible in the distributions of final eccentricities, which are more peaked at $e \sim 1$. Additionally, the semi-major axes distributions are skewed to large values, and clearly deviate from the simple power-law behaviour that we expect from the theory. 
However, the outcome distribution of the low-mass escapes (red lines) broadly agrees with theoretical expectations. We attribute this agreement to the fact that only 22\% of all the red escape trajectories are regular, despite their fraction over the total being 9\%. In contrast, regular trajectories account for about 42\% and 52\% of the trajectories for the green and blue escapes, respectively. In other words, not only does the total fraction of regular interactions increase for increasing mass, but also the relative fraction for each escaper mass.

After filtering out the regular islands by restricting the selection to simulations with $f_\mathrm{col}<2/3$, our numerical experiments align with theoretical expectations. Specifically, the eccentricity distributions converge towards a thermal distribution, consistent with theoretical predictions, while the distributions at larger semi-major axes exhibit the anticipated power-law behaviour. Deviations from theory persist at small semi-major axes due to limitations in number statistics.

Our findings have significant implications in the astrophysical application of statistical escape theories. Astrophysical predictions based on statistical theories will be inherently biased, because they exclude the outcomes of regular interactions. This issue is particularly concerning in the context of formation scenarios of gravitational wave mergers \citep[e.g.][]{kritos2022,mapelli2022,ginat2023}, where the final binary eccentricity, known to dramatically affect the gravitational wave radiation mechanism (at high $e$, the coalescence time $t_{\rm GW} \propto (1-e^2)^{7/2}$, see \citealp{peters1964}), is significantly different in the regular regions of the phase space. We delve deeper into this aspect in Section~\ref{sec:astro}.

\section{Chasing Laplace's demon}\label{sec:fractals}

\begin{figure*}
	\includegraphics[width=\linewidth]{{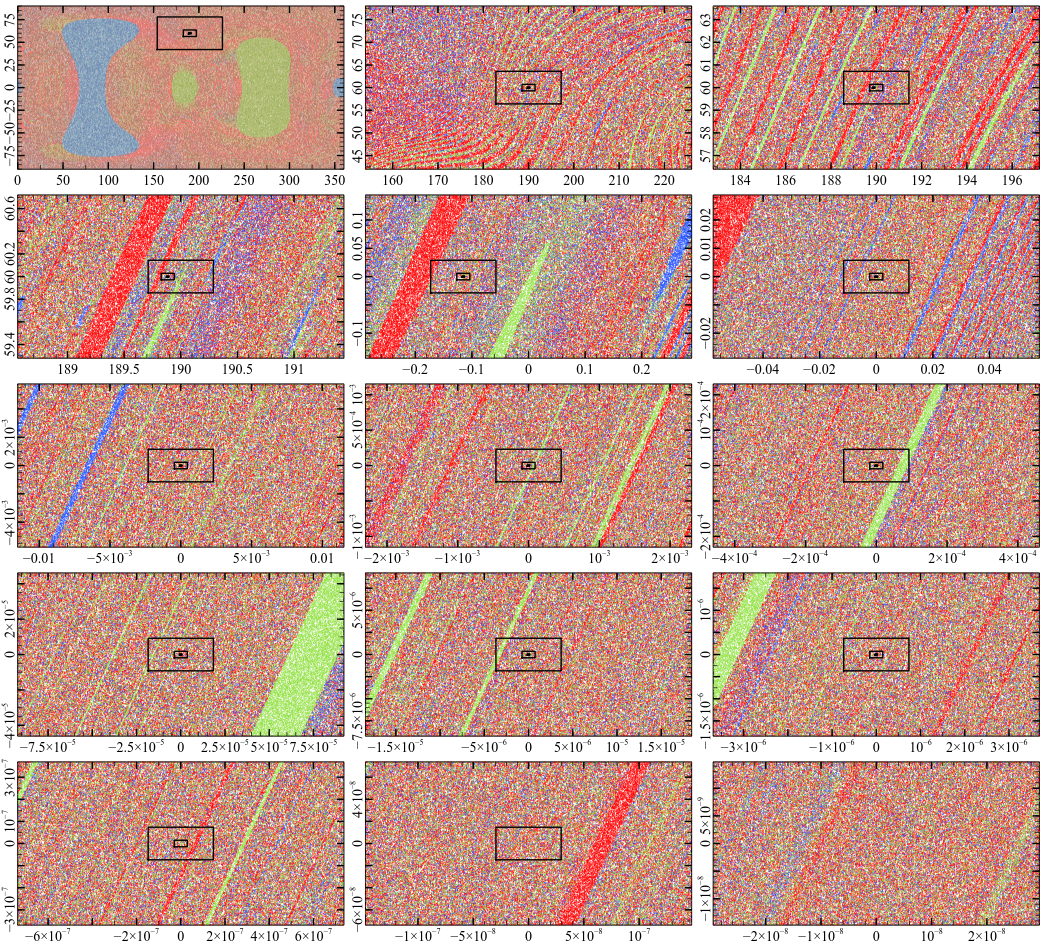}}
	
	\caption{Space of the initial configurations. The $x$-axis represents the phase, or true longitude $\lambda$ of the initial binary. The $y$-axis represents the inclination $\iota$ of the binary with respect to the line joining the binary centre-of-mass and the incoming single. In the first row, the absolute values are shown in degrees, while the other rows are shown as offset degrees from the central value. Each dot represents an initial configuration in $\lambda - i$ space, whose colour identifies the identity of the ejected particle. Red: $12.5 \,\rm M_\odot$. Blue: $15 \,\rm M_\odot$. Green: $17.5 \,\rm M_\odot$. Regions of sparse density may appear as white dots due to the white background. From left to right and top to bottom: each panel is a 5x5 zoomed-in version of the previous. The black boxes show the boundary box of the next three zoomed regions.}
	\label{fig:zoomphase_ejected}
\end{figure*}

In addition to the four large regular islands, Figure~\ref{fig:bigfig} highlights the presence of finer structures that look like narrow stripes. We explored whether there are regions between these structures without regular features, where all trajectories have outcomes uncorrelated with their neighbours. In Figure~\ref{fig:zoomphase_ejected} we show the initial phase space maps for the sequence of zoom-ins, beginning from a region of apparent chaos. As we zoom in, we observe the emergence of regular structures from what seemed at first glance to be a chaotic region. The regular structures are self-similar, taking the appearance of slanted stripes.

\begin{figure*}
	\includegraphics[width=\linewidth]{{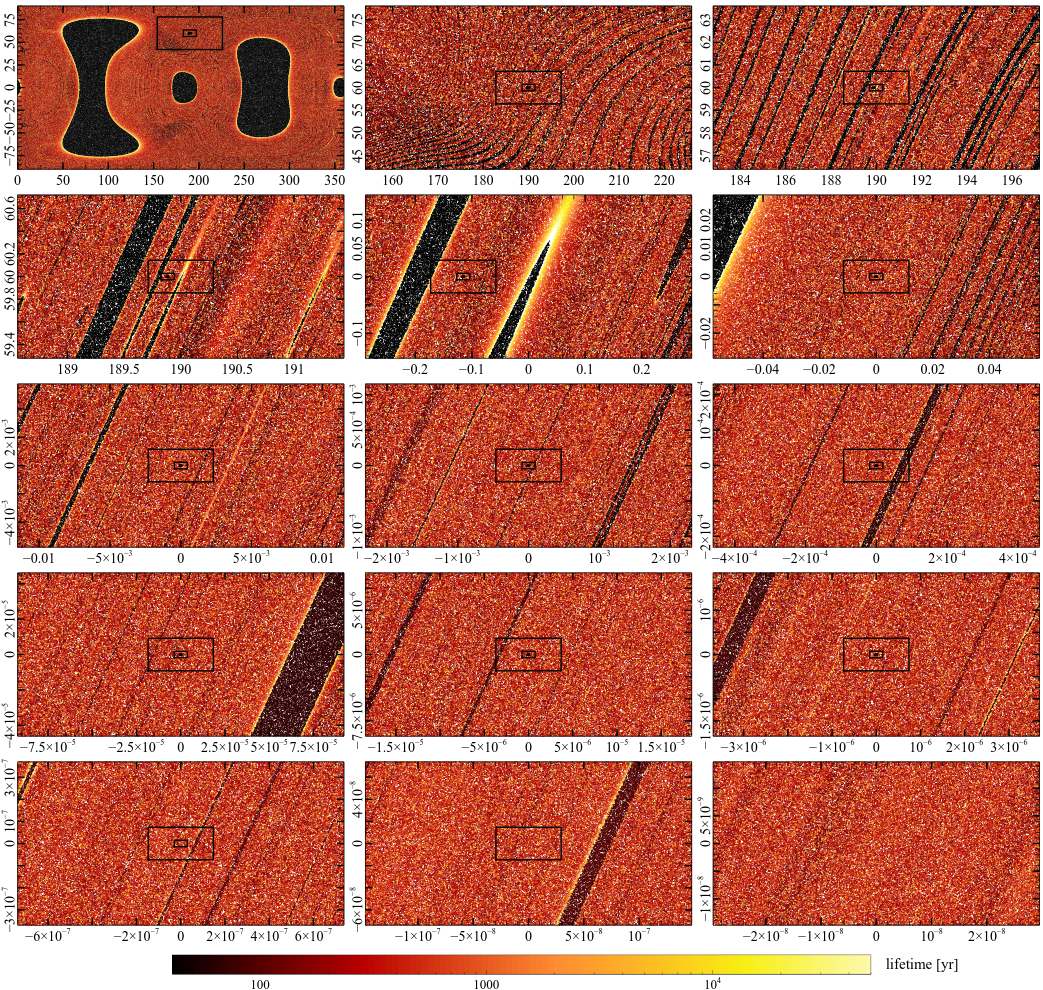}}
	
	\caption{Same as Figure~\ref{fig:zoomphase_ejected}, but colour coding the lifetime of the triples in logarithmic scale.}
	\label{fig:zoomphase_life}%
\end{figure*}

Similar patterns are visible if we colour code by the lifetime of the system, as in Figure~\ref{fig:zoomphase_life}. Uniform regions in the ejected colour map correspond to regions where the lifetime of the system is extremely short, comparable to the free fall time of the binary-single system $t_{\rm ff} =  \frac{\pi}{\sqrt{G(m_1+m_2+m_3)}} \left(\frac{d}{2}\right)^{3/2} \simeq 26 \,\rm yr$. Regular interactions have very short lifetimes because the system breaks up after a single close encounter between all three-particles, which is not enough to make the system highly chaotic. Several studies have indeed found that there is an approximate power-law relation between the Lyapunov time $t_{\rm Lyap}$ and the lifetime $t_{\rm life}$ of the gravitational 3BP \citep{lecar1992,orlov2010,mikkola2007,urminsky2009}. The main reason is that a triple system with a longer lifetime has more opportunities to undergo democratic resonances than a short-lived one. On the other hand, a long-lived system may also have spent most of its time in an excursion state, so the relation between $t_{\rm Lyap}$ and $t_{\rm life}$ is only approximate \citep{manwadkar2020}.

The short-lived, regular regions are contoured by edges of extremely long lifetimes, comparable to our integration time of $10^9 \rm\,yr$. These long lifetime regions arise from the presence of a smooth mapping within the regular regions, connecting the initial conditions to the outcome properties, including the escape velocity of the single particle. In other words, the outcome properties transition incrementally in terms of the left-over binary properties and those of the ejected single, with the same particle always being the one that is ejected. Eventually, in some direction in the initial phase space, the escape velocity of the single becomes smaller until the single becomes barely bound to the binary, and what was an ejection becomes a long-lived excursion. The interactions in this region can thus take an unbounded period of time to complete, because they spend most of their time in a temporary hierarchical triple state.

The self-similar nature of regular structures throughout the zoom-in is a manifestation of the fractal nature of chaos. Here we argue that chaos in the 3BP is multi-fractal in nature, meaning that different zoom levels and regions have different fractal dimensions.

\begin{figure*}
	 \sidecaption
	\includegraphics[width=12cm]{{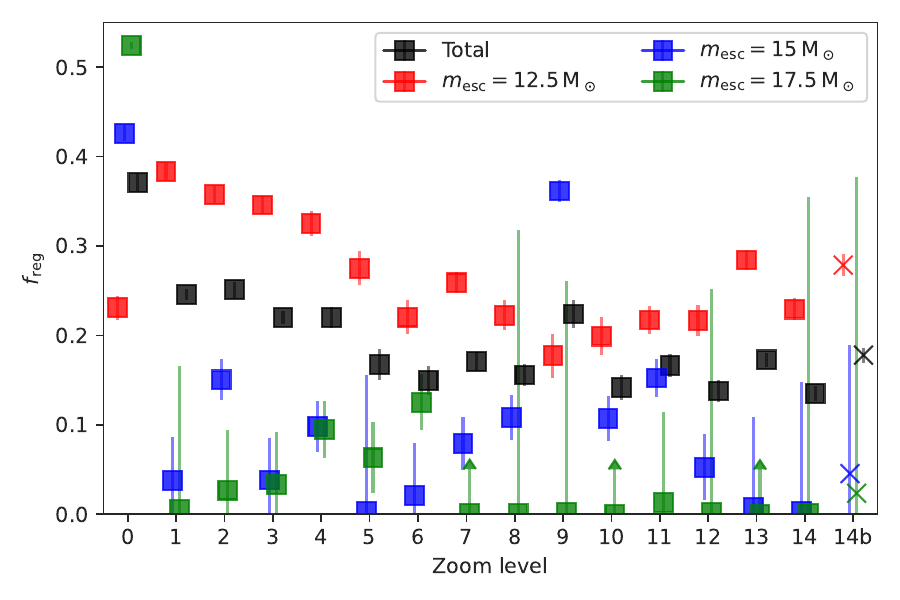}}
	\caption{Fraction of regular interactions ($f_{\rm col}>2/3$, $k=9$) for each zoom level. Zoom level 0 denotes the whole initial parameter space ($\iota \in [0,\pi]$, $\lambda \in [0,2\pi)$). Red, blue and green indicate the relative fraction (with respect to interactions of the same colour) of regular interactions where the $12.5$, $15$ or $17.5\,\msun$ particle escapes, respectively. The black line is the total fraction of regular interactions. The error range indicates the maximum between the Poissonian error and the difference in fraction estimate using $k=6$ and $k=12$ neighbours. The results for the 14th zoom-in are repeated using an arbitrary precision integrator, see Section~\ref{sec:numchaos}.
	}
	\label{fig:fcol}
\end{figure*}

First, from Figure~\ref{fig:fcol} we observe that the fraction of regular interactions $f_{\rm reg}$, as measured from our $k$-neighbours scheme, changes depending on which zoom level we are considering. Overall, the total percentage of regular interactions varies between 37\% and 15\% across the zoom levels. The relative fraction of regular trajectories for each escaper mass changes significantly over the zoom range. The low-mass escapes, initially having the lowest relative fraction of regular trajectories in the outer box, display the highest relative fraction in all zoom levels except the 9th one. This discrepancy is evident from Figure~\ref{fig:zoomphase_ejected}, where the 9th zoom-in prominently features a large green stripe, constituting the majority of the regular space.
Conversely, the high-mass escapes, initially exhibiting the highest fraction of regular trajectories in the outer box, display the least relative fraction in most of the zoom levels. This is primarily due to the paucity of high-mass escapes in the zoom-ins, compounded by our numerical resolution limitations as detailed in Section~\ref{sec:numchaos}. The varying fraction of regular interactions from the outer box to the zoom-ins underscores the qualitative difference in the system behaviour across different scales.

To better estimate the fractal nature of the phase space maps, we can measure the fractal dimension of every zoom-in. One of the most common ways to estimate the fractal dimension of a set of points is to calculate the box-counting dimension, or Minkowski–Bouligand dimension \citep{schroeder1991}. Unfortunately, this technique is not suited for our simulations, because we are limited by resolution at both the small scales (set by the average distance between realization in a box) and large scales (set by the outer boundary of each zoom-in box).

Therefore, we adopt instead the 2-point correlation dimension, which is known to be accurate even when only a small number of points is available \citep{grassberger1983}. This technique works by counting the number of pairs $C(l)$ whose distance is less then $l$. Then, for small $l$,  $C(l)$ grows as a power law, that is $C(l) \propto l^{D_2}$, where $D_2$ is the two-point correlation dimension, which is guaranteed to be close (strictly speaking, equal to or less than) the fractal dimension of the set. \REV{In a finite box, points near the edges have fewer neighbours within a distance $l$ compared to points well inside the box. This reduction in neighbours leads to an underestimation of $C(l)$ near the edges. To address this, we applied an edge-correction factor to $C(l)$, estimated by comparing $C(l)$ from a uniform distribution with and without a toroidal topology at the boundaries}.

\REV{
The fractal dimension is a measure of the complexity of a fractal surface embedded in a space of dimension $D$, indicating how it scales with size. It has found applications in several astrophysical scenarios, for instance, to measure the clustering degree in clusters of proto-stars \citep{larson1995,cartwright2004,ballone2021}. In our case, the initial phase space of our 3BP setup resides in $D=2$. A fractal dimension of $D_2 = D = 2$ implies that the considered set is not fractal but smooth. This suggests that the underlying dynamics is regular rather than chaotic, as the set uniformly populates the embedding space.
This is clear when examining the large islands in Figure~\ref{fig:bigfig}, which are caused by the presence of a smooth mapping between initial and final phase spaces, a characteristic of regular dynamics. The lower the value of $D_2$, the sparser the set becomes, meaning it fills less of the embedding space. It is also possible that the correlation function $C(l)$ may not be adequately described by a single power-law, suggesting that the fractal exhibits varying levels of complexity or self-similarity across different scales. In such instances, the fractal is more accurately characterized by a spectrum of fractal dimensions, giving rise to what is known as a multi-fractal.
}

\begin{figure*}[h]
	\centering
	\includegraphics[width=1.0\textwidth]{{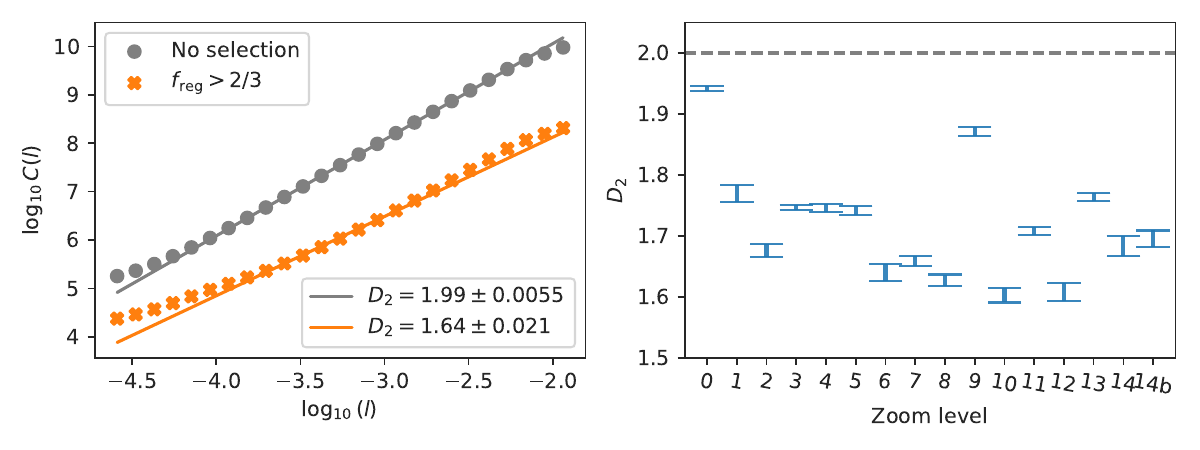}}
	\caption{Left panel: measure of the two-point correlation dimension of the regular interactions of the sixth zoom-in level. The ordinate $C(l)$ is the number of pairs with phase-space distance less than scale $l$. The grey dots includes all the realizations in the simulations, whose power-law fit (grey line) is close to its expected dimension ($D \simeq 2$). The orange dots only consider the regular set ($f_{\rm col} > 2/3$), and its fractal dimension is lower than 2. Right panel: fractal dimension for each zoom level, as measured from the 2-point correlation dimension. The error bars represent the least-squared fit. The results for the 14th zoom-in are repeated using an arbitrary precision integrator, see Section~\ref{sec:numchaos}.
	}
	\label{fig:combofractal}
\end{figure*}

\REV{As shown in the left-panel of Figure~\ref{fig:combofractal}, measuring $D_2$ for uniformly distributed points in a box yields a dimension consistent with the expected one, that is $D_2 \simeq 2$. It is important to note how the $C(l)$ curve deviates from a power-law at the extrema, where we either approach the resolution of our set or the large-scale boundary of the box. To prevent these boundaries from affecting our least squared fit to the power-law exponent, we weight the data using a Gaussian function centred on the mean scale.}

We then select only the regular set, here again defined as the ensemble of realizations with a same-colour fraction $f_\mathrm{col}> 2/3$ among the $k=9$ nearest neighbours, to select only the most uniform region. We find a fractal dimension that is smaller than 2 at all zoom levels, which points out the fractal nature of regular trajectories in the phase space, because it is fractional and smaller than the dimensions of the embedding space. However, each zoom-level presents a different fractal index, as seen in the right panel of Figure~\ref{fig:combofractal}. Overall, the fractal dimension $D_2$ changes across the zoom-ins, which confirms our hypothesis of the multi-fractal nature of chaos in the 3BP. Interestingly, $D_2$ remains constant across the 3rd, 4th, and 5th zoom-ins, which also exhibit remarkably similar features in Figure~\ref{fig:zoomphase_life} (thick regular stripes). Similarly, the 6th, 7th, and 8th zoom-ins, characterized by thin regular stripes, also display comparable fractal dimensions. \REV{The fractal dimension for zoom levels 6,7,8 is ${\sim}0.1$ smaller than that of zoom levels 3,4,5, which are consistent with having sparser, thinner stripes.} The most significant change in $D_2$ occurs with the 9th zoom-in, featuring a large regular stripe, resulting in higher $D_2$ values, indicative of a more two-dimensional structure. \REV{The fractal dimension of the full phase space (zoom level 0) is also consistent with what we observe in Figure~\ref{fig:bigfig}. At this scale and resolution, the majority of the regular phase space is contained within the four large islands, which lack substructures, so the fractal dimension is closer to 2. 
An analogous behaviour can be observed in the alternative zoom-ins of  Appendix~\ref{sec:extrazooms}, which targeted the border of the green regular region in the 4th zoom-in.
}

\section{The impact of numerical chaos}\label{sec:numchaos}
\begin{figure*}
	\sidecaption
	\includegraphics[width=0.7\textwidth]{{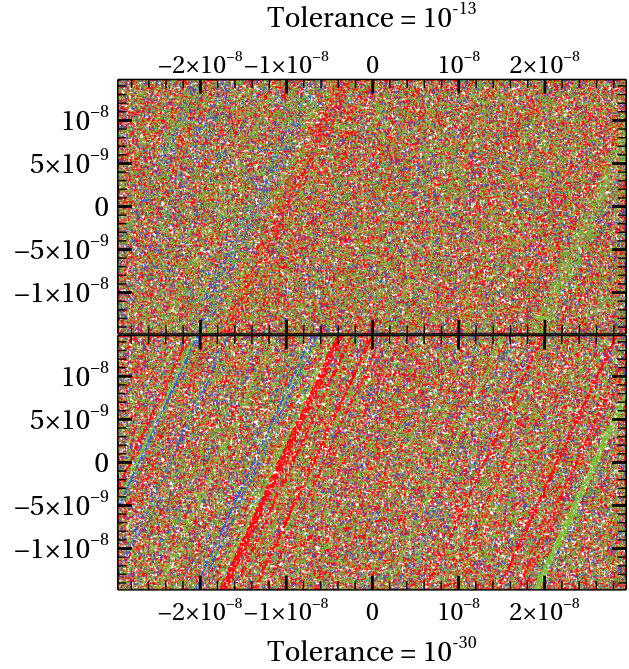}}
	\caption{Top panel: last zoom-in set from Figure~\ref{fig:zoomphase_ejected} (zoom level 14), run with the \textsc{tsunami} code with a tolerance of $10^{-13}$. Bottom panel: same set, run with the \textsc{brutus} code with a tolerance of $10^{-30}$.
	}
	\label{fig:numchaos}
\end{figure*}

In general, numerical integrations are constrained by the inherent limitations of floating-point arithmetic precision, the temporal discretization of the solution, and the truncation error of the integration algorithm (see \citealp{portegieszwart2018} for a discussion). The simulations presented thus far also suffer from these limitations (even though in \textsc{tsunami} the truncation errors are compensated by employing Richardson extrapolation, \citealp{numrec}). Small integration errors accumulate over time and result in a deviation of our approximate solution from the mathematical one (i.e. the true mathematical solution that is guaranteed to exist by virtue of the Cauchy–Lipschitz theorem), indicating a lack of numerical convergence. Previous studies have highlighted that as long as numerical errors are below a certain threshold (impossible to determine a priori), an ensemble of approximate solutions can be statistically indistinguishable, even when individual solutions may not all be numerically converged \citep{boekholt2015}.

Chaos influences the required accuracy and precision for simulations to achieve numerical convergence. In chaotic systems, small errors propagate and amplify much faster than in regular systems. Consequently, regular systems have less stringent numerical requirements for obtaining a converged solution \citep{portegieszwart2018}. The degree of chaos in a self-gravitating system depends on various factors. Generally, close encounters make a system more chaotic \citep{suto1991,portegieszwart2023,boekholt2023,boekholt2024a}, and systems with a larger number of particles have a shorter Lyapunov timescale \citep{heggie1988,kandrup1991,kandrup1992a,kandrup1992b,goodman1993,kandrup1994,hemsendorf2002,dicintio2019,dicintio2020,portegieszwart2022}. In the 3BP, chaos arises from the close encounters during democratic resonances.

Here, we have the opportunity to quantify how numerical errors manifest in the initial phase space, in addition to understanding their impact on the statistical outcome of the 3BP. We anticipate that the effect of numerical errors will be more pronounced in the set with zoom level 14, the last set of Figure~\ref{fig:zoomphase_ejected}, where the separation of neighbouring initial conditions in the phase space is the smallest.

To investigate this, we rerun the simulations using the $\textsc{brutus}$ code \citep{boekholt2015}, which employs arbitrary-precision arithmetic. We set the error tolerance parameter to 10$^{-30}$, which is more stringent by 17 orders of magnitude compared to \textsc{tsunami}, and a 256 bits floating point arithmetic, in place of the commonly used double-precision (64 bits).

In Figure~\ref{fig:numchaos}, we compare the phase space of outcome states between the two integrators. Owing to its enhanced accuracy and precision, $\textsc{brutus}$ resolves regular regions with higher resolution. The impact of numerical errors is evident in artificially mixing the phase space, diminishing the coherence and definition of regular regions. Fine regular stripes resolved in the bottom panel lose coherence at lower accuracy and blend together with neighboring stripes. We term this artificial mixing as `numerical chaos'. 

Because the term numerical chaos has only been introduced once before in the literature \citep{numchaos}, it is essential to provide a clear definition to prevent confusion. Numerical chaos arises when numerical errors are amplified within a system characterized by both regular and chaotic regions in its phase space. For numerical chaos to occur, chaotic regions must exist, which diffuse throughout the phase space and infiltrate otherwise regular regions because of numerical errors. Put simply, numerical errors cannot induce chaos in a system that is entirely regular.

Consistent with the findings of \citet{boekholt2015}, numerical chaos does not significantly affect the macroscopic outcomes of the interactions, with the difference in outcome statistics being less than 0.5\% (Table~\ref{tab:numchaos1}). However, it is important to note that for this statement to hold true, the sample size must be sufficiently large.

However, it notably reduces the number of regular interactions, decreasing them by 25\% overall, and by close to 80\% for the rare escapes involving the more massive particle (green, Table~\ref{tab:numchaos2}). This decrease in the regular fraction of phase space introduces noise when comparing outcome properties with the statistical theories, as it includes regular trajectories misclassified as chaotic, which would have otherwise been excluded. This result is also visualized in Figure~\ref{fig:fcol}, where the last zoom-in (14b) is the one run with $\textsc{brutus}$. Despite the loss of coherence in the regular stripes due to numerical errors, the measured fractal dimension remains unaffected (see 14b in Figure~\ref{fig:combofractal}).

These results indicate that a small source of numerical noise has the same effect as a physical perturbation, which enables transport between neighbouring Hamiltonian flows \citep{kandrup2004}. Numerical chaos can therefore accelerate the rate at which an ensemble of orbits diffuse in the phase space, hiding the intrinsic regularity of a dynamical system.


\begin{table}
	\centering
	\caption{Fraction of outcomes for the last zoom-in of Figure~\ref{fig:zoomphase_ejected}, for the two different integrators. \vspace{3pt}}\label{tab:numchaos1}
	\begin{tabular}{c|ccc}
  &  $f_{\rm red}$  &  $f_{\rm blue}$  &  $f_{\rm green}$ \\\hline\hline
\textsc{brutus}  &  0.5824  &  0.2644  &  0.1532  \\
\textsc{tsunami}  &  0.5826  &  0.265  &  0.1525  \\
$\Delta$  &  $+0.0219$\%  &  $+0.223$\%  &  $-0.467$\%  \\
	\end{tabular}
	\tablefoot{Columns from left to right: fraction of outcomes with escaper of mass $12.5$, $15$ or $17.5\,\msun$, respectively. The bottom row shows the relative difference in percentage.}
\end{table}

\begin{table*}
	\centering
	\caption{Fraction of regular interactions for the last zoom-in of Figure~\ref{fig:zoomphase_ejected}, for the two different integrators.}\label{tab:numchaos2}
	\begin{tabular}{c|cccc}
 &  $f_{\rm reg,red}$  &  $f_{\rm reg,blue}$  &  $f_{\rm reg,green}$  &  $f_{\rm reg}$ \\\hline\hline
\textsc{brutus}  &  0.16$\pm$0.00089  &  0.013$\pm$0.0022  &  0.0037$\pm$0.0012  &  0.18$\pm$0.0043  \\
\textsc{tsunami}  &  0.13$\pm$0.0017  &  0.0013$\pm$0.00076  &  0.00021$\pm$0.00016  &  0.13$\pm$0.0026  \\
$\Delta$  &  $-18.2$\%  &  $-89.4$\%  &  $-94.5$\%  &  $-24.9$\%  \\
	\end{tabular}
	\tablefoot{Columns from left to right: fraction of outcomes with escaper of mass $12.5$, $15$ or $17.5\,\msun$, respectively. The last column indicates all regular interactions. The bottom row shows the relative difference in percentage.}
\end{table*}

\section{Perturbations below the Planck scale}\label{sec:planck}

\begin{figure*}
	\centering
	\includegraphics[width=\linewidth]{{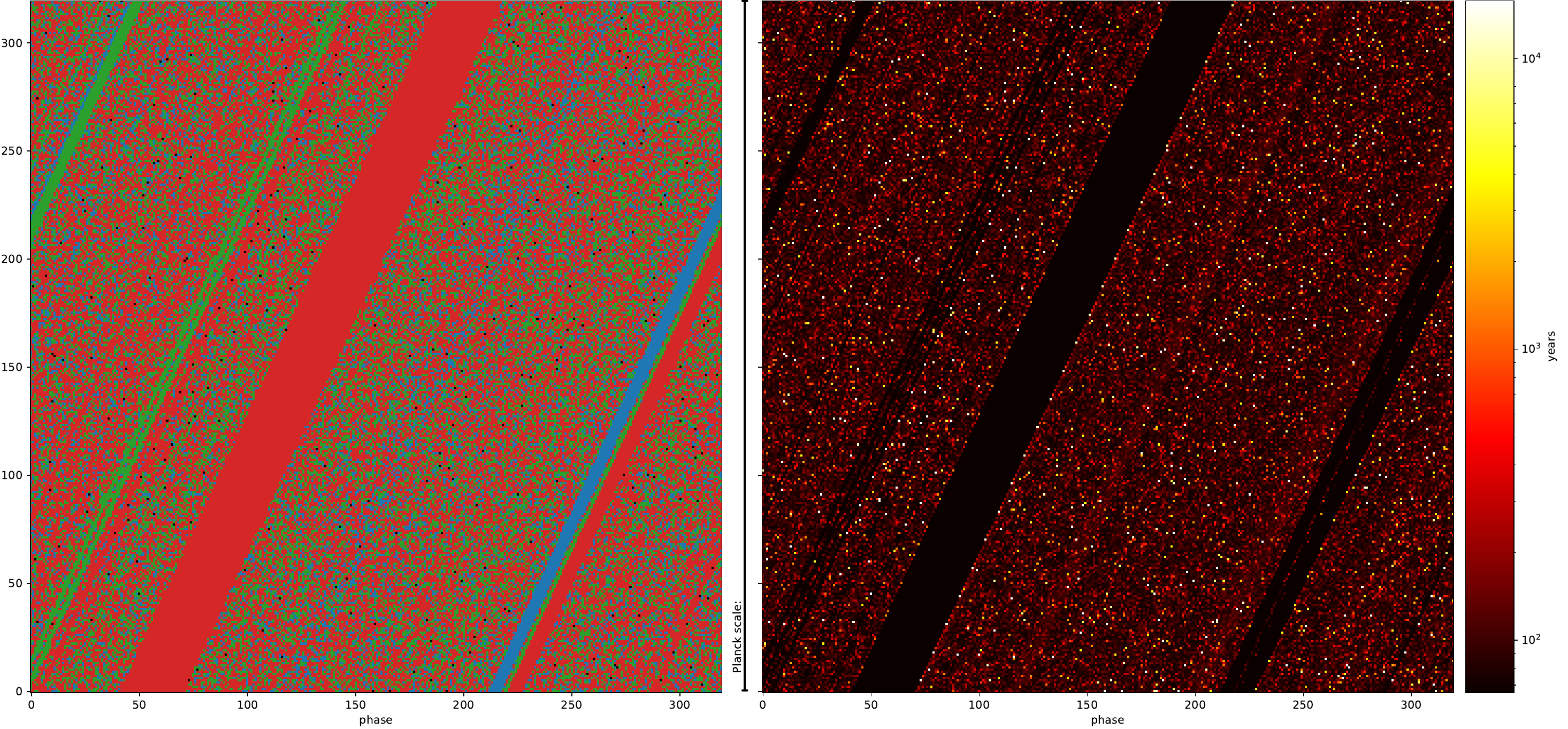}}
	\caption{Space of the initial configurations, colour-coded by the ejected particle (left) and lifetime of the system (right). Each pixel corresponds to a realization on a $320\times320$ grid. As in Figure~\ref{fig:zoomphase_ejected}, the axes denote binary phase and inclination, but since displaying the actual values in degrees would require too many digits, we just show the realization number on the grid. The size of the box is equivalent to a perturbation in the particles's positions comparable to the Planck length. The simulations were run with \textsc{brutus} with a tolerance of $10^{-63}$ and a word length of 356 bits.
	}
	\label{fig:planckscale}
\end{figure*}

Employing an arbitrarily accurate integrator allows us to unlimitedly zoom into the phase space. Having demonstrated the fractal nature of the regular structures, we expect the self-similar patterns of regular and chaotic regions to continue down to any scale. 
From zoom level 14, our last zoom-in, we perform an additional zoom-in of ${\approx} 2 \times 10^{41}$ in each dimension and run 102400 realizations using \textsc{brutus}. The result is displayed in Figure~\ref{fig:planckscale}.

At this scale, each pixel differs from the neighbouring one by $5\times 10^{-46}$ radians. Converting the difference in inclination and phase to difference in particle position, we estimate that the size of the box has a length comparable to the Planck length ($l_{\rm P} = 1.0804 \times 10^{-46} \rm\,au$), and the particles in neighbouring configurations are distant from each other by a mere $l_{\rm P}/320$. 

\REV{With compelling evidence demonstrating that chaos and regularity persist at any scale, including scales below the Planck length, any microscopic stochastic effect below the Planck length has the potential to propagate macroscopically \citep[see also][]{boekholt2020}. While chaos washes away any microscopic effect by hiding the correlation between initial conditions and final outcomes, regular behaviour offers a means to connect micro- and macroscopic physics. Specifically, any potential microscopic perturbation to particle positions will not produce a different macroscopic outcome if the initial position resides within a regular region, as opposed to residing within a chaotic region.}

\section{Astrophysical implications}\label{sec:astro}

Statistical escape theories of the 3BP have recently gained traction in astrophysics due to their computational efficiency compared to direct three-body simulations. Notably, these theories have found various applications in predicting the properties of merging black holes resulting from dynamical interactions in dense stellar environments \citep[e.g.,][]{kritos2022,mapelli2022,ginat2023}. Additionally, a recent study has utilized statistical escape theories to model the ejection of extra-tidal stars from globular clusters in the Galactic halo \citep{groundin2023}.

Our findings raise concerns regarding the application of statistical escape theories to astrophysical scenarios. While all trajectories we calculated numerically are physically plausible and can occur in nature, statistical escape theories only predict those trajectories that exhibit sufficient chaotic behaviour to mix the phase space. In our setup, this results in the exclusion of approximately 37\% of the initial phase space. As demonstrated in Section~\ref{sec:stats}, this exclusion biases the distribution of outcome parameters (such as binary semimajor axis and eccentricity, and single escape velocity), as well as the escape probability of the single, based on its mass.

\begin{figure}
	\centering
	\includegraphics[width=1\linewidth]{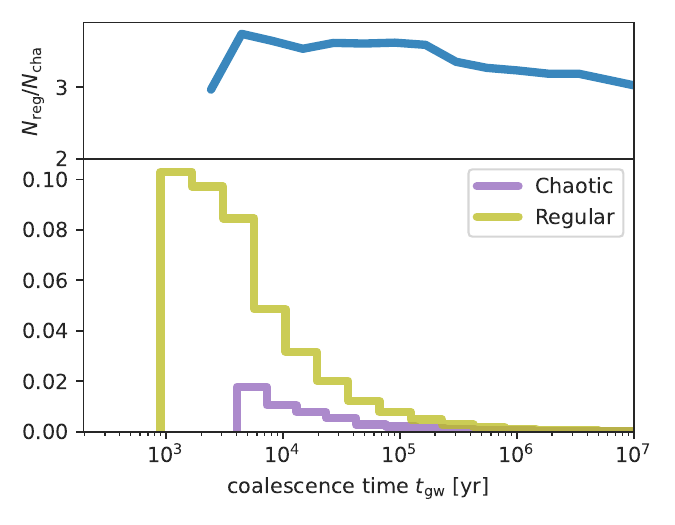}
	\caption{Top panel: ratio of the cumulative number of binaries that formed from regular interactions to that of binaries that formed from chaotic interactions, as a function of gravitational wave coalescence time $t_{\rm gw}$.
	Bottom panel: coalescence time distribution of final binaries for the regular ($f_{\rm col} \leq 2/3$, olive) and chaotic ($f_{\rm col} > 2/3$, purple) interactions. Both distributions are normalized to the total sample.
	}
	\label{fig:coaltime}
\end{figure}

In Figure~\ref{fig:coaltime}, we illustrate how these biases impact the estimation of the gravitational wave coalescence time $t_{\rm gw}$ for the final binaries in our simulations, assuming they are black holes \citep{peters1964}. Despite constituting only about a third of the total, binaries formed from regular interactions exhibit significantly shorter coalescence times compared to those formed from chaotic interactions, because of their higher eccentricity. Moreover, binaries resulting from regular interactions with short ($t_{\rm gw} < 10^7\rm\,yr$) coalescence times are roughly three times as numerous as those from chaotic interactions.
Our analysis therefore suggests that statistical escape theories overestimate the coalescence time, and thus underestimate the merger efficiency of binaries formed from three-body interactions. Another likely consequence of the bias in eccentricity is the underestimation of the fraction of eccentric mergers that may be detectable with gravitational wave interferometers \citep{2019ApJ...883..149A,2021ApJ...921L..31R,2022NatAs...6..344G,2023arXiv230803822T}.

We emphasize that the fraction of regular trajectories, as well as their outcome properties, should vary depending on the specific initial setup of the 3BP (see Appendix~\ref{sec:extrasims}). Nevertheless, we anticipate the coexistence of regular and chaotic regions to be a natural feature of the 3BP, as supported by findings in other studies with different setups \citep[e.g.][]{manwadkar2021,parischewsky2023,manwadkar2023}. For example, \citet{manwadkar2023} investigated the `chaotic absorptivity' (the probability for an interaction to enter chaotic motion) for hyperbolic binary-single encounters with varying initial binary angular momentum and energy. Their figure~5 illustrates the persistent separation between chaotic and regular motion in the energy-angular momentum space, with the fraction of chaotic phase space decreasing for larger initial binary separations. Thus, it is important to evaluate the biases introduced by the use of statistical escape theories to model astrophysical scenarios on a case-by-case basis. A comprehensive assessment of these biases is beyond the scope of this paper.

\REV{
Another caveat is that our simulations are fully Newtonian and do not account for collisions between finite-sized particles. The inclusion of post-Newtonian (PN) corrections would introduce the possibility of mergers (or collisions), driven by radiation reaction terms (2.5 and 3.5 PN order corrections) and apsidal precession effects (1, 2, and 3 PN order corrections). 
However, the presence of dissipation terms is unlikely to significantly affect the mixing of chaos and regular regions. This is because the radiation reaction terms depend steeply on the distance between two particles, and are efficient only during the closest approaches. Previous studies have shown that radiation terms create narrow regular regions in the phase space, corresponding to collision outcomes, while leaving the rest of the phase space largely untouched \citep{samsing2019,parischewsky2023}. On the other hand, precession terms have been found to reduce chaos in $N$-body systems, but these effects are only significant in the strong gravity regime ($v/c \gtrsim 0.005$, \citealt{portegieszwart2022}).

In summary, the inclusion of PN corrections would likely aggravate the bias of statistical escape theories by increasing the regularity of the system. We defer a complete numerical verification of this statement to future work, as we have disabled PN corrections for the present study to enable a direct comparison of our simulations with statistical escape theories, which assume Newtonian gravity.
}

\section{Conclusions}\label{sec:conclus}

Our numerical investigation highlights the mixed nature of chaos in the gravitational 3BP. At all scales, chaos is intermixed with regular regions where the triple breaks up very quickly, exhibiting regular behaviour, identified as a smooth mapping between initial conditions and outcome properties (Figure~\ref{fig:bigfig}). These regular interactions comprise about 37\% of the total phase space.

Statistical escape theories accurately reproduce the numerical experiments in the chaotic regions of the phase space, where orbits are fully mixed and are close to a microcanonical population. In contrast, the statistical outcomes of regular interactions defy the predictions of escape theories. In particular, the outcome distribution of binary eccentricities and semi-major axes exhibit strong deviations from the theoretical predictions (Figure~\ref{fig:comparedist}).

Regular regions have edges with extremely long interactions. These correspond to systems whose lifetime goes asymptotically to infinity due to the regular mapping between initial conditions and the energy of the escaping single during the first interaction. In between the escape of the single and its retention, there exist an infinite number of orbits with infinitesimally small binding energy (Figure~\ref{fig:zoomphase_life}).

However deep we zoom in on the initial phase space, we find regular and chaotic regions appearing in a self-similar pattern. The number of regular islands increases with the resolution in our phase space, indicating that the mixing between chaotic and regular regions is fractal in nature (Figure~\ref{fig:zoomphase_ejected}). We measured its fractal dimension at various zoom levels of the phase space. No single exponent can fully describe the fractal dimension of the phase space, highlighting the multi-fractal nature of chaos in the 3BP (Figure~\ref{fig:combofractal}).

Eventually, the chaotic behaviour reaches the quantum realm, where a difference in positions below the Planck length results in macroscopic different outcomes (Figure~\ref{fig:planckscale}). The mixing of regular and chaotic regions persists below this scale, coupling macro and microscopic physics.

Numerical errors induce a spurious mixing of the phase space, disrupting otherwise regular regions (Figure~\ref{fig:numchaos}). We term this as numerical chaos. While individual simulations are unreliable due to numerical errors, the statistical correctness of the ensemble of simulations is preserved to less than one percent error. However, numerical chaos introduces a mixing of the phase space that leads to an underestimation of the regular phase space by ${\sim}$25\% (Figure~\ref{fig:fcol}). Misclassifying regular trajectories as chaotic introduces noise when comparing the results of numerical experiments with statistical escape theories, diminishing the apparent accuracy of the statistical theories.

One potential critique of our work could be that our results are only applicable to our specific three-body setup, which admittedly represents a small portion of the vast parameter space of the 3BP. To address this concern, we conducted additional numerical experiments involving hyperbolic binary-single scatterings with an initial velocity at infinity and impact parameter \REV{(see Appendix~\ref{sec:extrasims}). Remarkably, we observe regular patterns similar to those in Figure~\ref{fig:bigfig}, and estimate the percentage of regular interactions to range from 28\% to 84\%, depending on the initial setup (Figure~\ref{fig:extrasim}). These complementary results indicate the robustness of our findings.} Moreover, our conclusions are supported by the results of \citet{manwadkar2023}, who found that chaotic and regular motion occupy distinct regions in the energy-angular momentum space of hyperbolic binary-single encounters, even when statistically averaged over the binary phase.

Another crucial variation of parameters that we have not yet explored is the masses of the three bodies, or rather, their mass ratio. Intuitively, as the mass of the third body decreases, the motion of the two most massive bodies is expected to become increasingly regular. In the extreme regime of the restricted 3BP with $m_3 \ll m_1, m_2$, the binary moves unperturbed on a Keplerian orbit, while the motion of the test particle remains chaotic. \citet{manwadkar2020} showed that as the mass contrast between the most and least massive bodies increases, the regular islands depicted in Figure~\ref{fig:bigfig} begin to disintegrate. Exploring the transition between the equal mass and test particle limits, as well as the limitations of escape theories in this context, presents an interesting direction for future research.

Our findings underscore that regular regions in the phase space pose challenges to the effectiveness of statistical escape theories as predictive tools for three-body interactions. Consequently, caution is advised when applying such tools to astrophysical problems \citep[e.g.][]{kritos2022,mapelli2022,groundin2023,ginat2023}. Specifically, the eccentricity distribution of final binaries from regular trajectories diverges significantly from the expectations of statistical escape theories (Figure~\ref{fig:coaltime}), which introduces a bias into gravitational wave population synthesis models reliant on such theories.




\begin{acknowledgements}
The authors thank the editor Beno\^it Noyelles and the anonymous referee for suggestions that improved the paper. AAT acknowledges support from JSPS KAKENHI Grant Number 21K13914 and from the European Union’s Horizon 2020 and Horizon Europe research and innovation programs under the Marie Sk\l{}odowska-Curie grant agreements no. 847523 and 101103134. All the simulations were performed on the \texttt{awamori} computer cluster at The University of Tokyo. TCNB is supported by an appointment to the NASA Postdoctoral Program at the NASA Ames Research Center, administered by Oak Ridge Associated Universities under contract with NASA.
NWCL gratefully acknowledges the generous support of a Fondecyt General grant 1230082, as well as support from Núcleo Milenio NCN2023\_002 (TITANs) and funding via the BASAL Centro de Excelencia en Astrofisica y Tecnologias Afines (CATA) grant PFB-06/2007.  NWCL also thanks support from ANID BASAL project ACE210002 and ANID BASAL projects ACE210002 and FB210003. AAT, SPZ and TCNB also thank Anna Lisa Varri for organizing the ``Chaotic rendezvous'' meeting at the University of Edinburgh, where part of this work was completed. AAT would like to express gratitude to those who contributed with insightful discussions at different stages of this work, including Alessandro Ballone, Rosemary Mardling, Barak Kol, Mario Pasquato.
\end{acknowledgements}

\bibliographystyle{aa}

\bibliography{totalms,chaos} 

\begin{appendix}
	\section{Simulations with hyperbolic setup} \label{sec:extrasims}
\begin{figure*}
		\centering
		\includegraphics[width=1\linewidth]{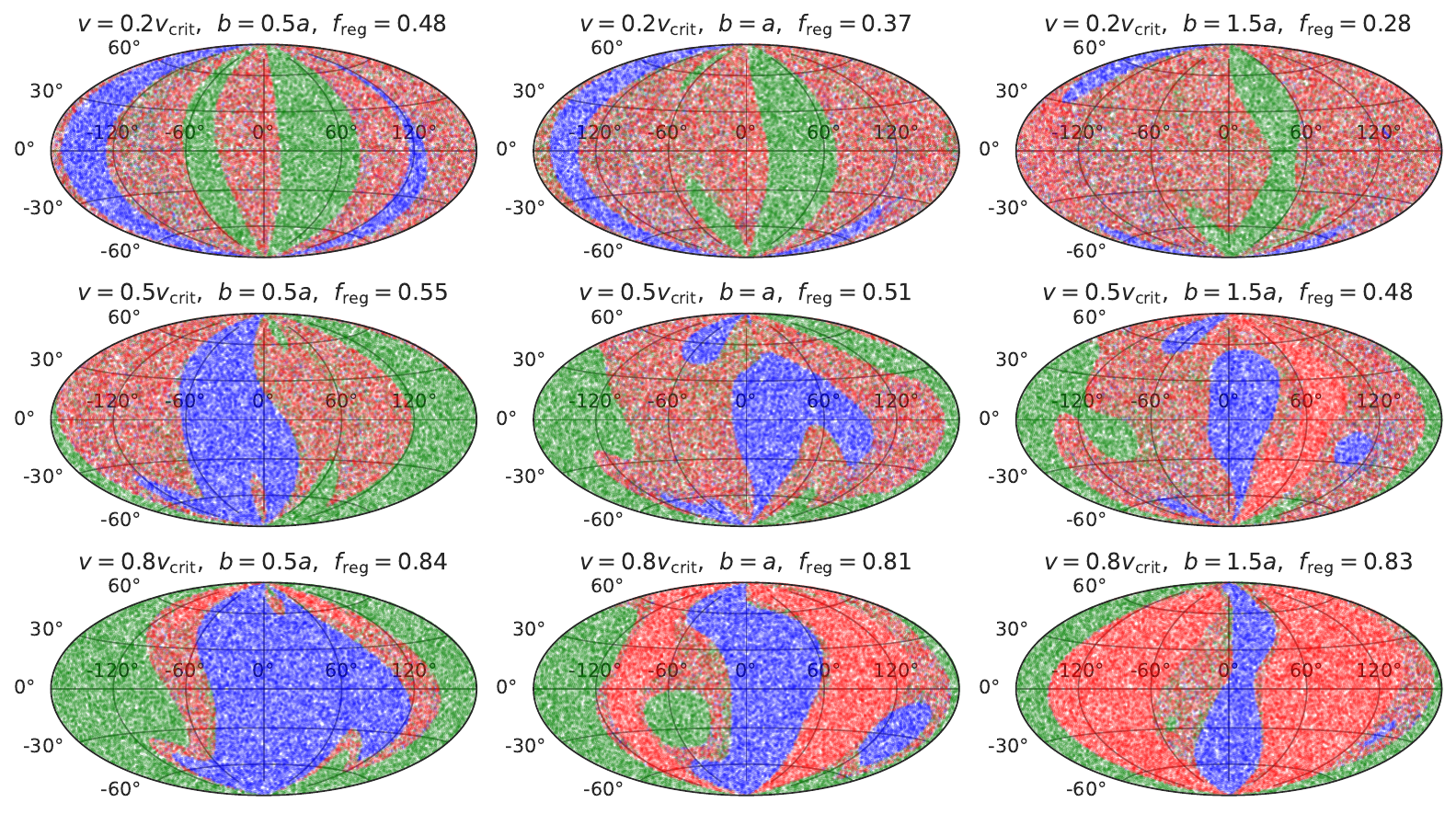}
		\caption{Initial condition space in inclination $\iota$ ($y$-axis) and true longitude $\lambda$ ($x$-axis), in Hammer projection. In every panel, each dot represents an individual realization among $5\times10^4$. The dots are colour-coded by the identity of the escaping particles. Red: $12.5 \,\rm M_\odot$. Blue: $15 \,\rm M_\odot$. Green: $17.5 \,\rm M_\odot$. In these sets, the binary-single encounters are on hyperbolic orbits with impact parameter $b$ and velocity at infinity $v$. From top to bottom: $v / v_\mathrm{crit} = 0.2, 0.5$ and $0.8$ (see Equation~\ref{eq:vcrit} for the definition of $v_\mathrm{crit}$). From left to right: $b / a = 0.5, 1$ and $1.5$. The fraction of regular interactions ($f_\mathrm{reg}$) is displayed on the upper-left of the projection.
		} 
		\label{fig:extrasim}
	\end{figure*}

	\REV{
		We performed additional simulations to investigate whether the mixed chaotic-regular phase space observed in the 3BP is a characteristic feature of our specific initial setup or not. In these new nine sets of simulations, we initialized the binary and the single on hyperbolic trajectories with a velocity at infinity $v$ and impact parameter $b$. For each set, we run $5\times10^4$ realizations, varying the binary phase and inclination of the incoming single, but fixing velocity at infinity and impact parameter. Specifically, we choose a combination of $v / v_\mathrm{crit} = 0.2, 0.5$, and $0.8$, where $v_\mathrm{crit}$ is the critical velocity for which the total energy of the three-body system vanishes:
		\begin{equation}\label{eq:vcrit}
			v^2_\mathrm{crit} = \frac{G m_1 m_2}{a} \,\frac{m_1 + m_2 + m_3}{m_3 (m_1 + m_2)} \,,
		\end{equation}
		and $b / a = 0.5, 1$, and $1.5$. The masses ($m_1$, $m_2$, $m_3$) and semi-major axis ($a$) of the binary are kept the same as in our fiducial experiment. We run $5\times 10^4$ realizations for each set.
		
		Figure~\ref{fig:extrasim} shows the initial phase-space maps, akin to Figure~\ref{fig:bigfig}, for each of the nine sets. Despite the radically different initial setups, we observe similar patterns to those in our fiducial setup, where the centre-of-mass velocities of the binary and the single are zero.
		
		We calculated the fraction of regular interactions $f_\mathrm{reg}$ as in Section~\ref{sec:stats}, employing $k=9$ neighbours. These fractions are depicted in Figure~\ref{fig:extrasim}. Overall, the fraction of regular interactions ranges from $0.28$ to $0.84$, and for small $v$ and $b$ (top-left panel) it is comparable to our fiducial setup. Notably, for $v/v_\mathrm{crit} = 0.2$ and $b / a = 0.5$, the phase space closely resembles the one in Figure~\ref{fig:bigfig}. However, as $b$ and $v$ increase, the phase-space maps begin to morph, displaying stronger asymmetries and more complex shapes. The fraction of regular interactions increases with $v$, most likely due to the increase in flybys. Curiously, the fraction of regular interactions decreases for increasing impact parameter at $v/v_\mathrm{crit} = 0.2$.
	}
	
	\section{Zoom-ins in other regions} \label{sec:extrazooms}
\begin{figure*}
		\centering
		\includegraphics[width=1\linewidth]{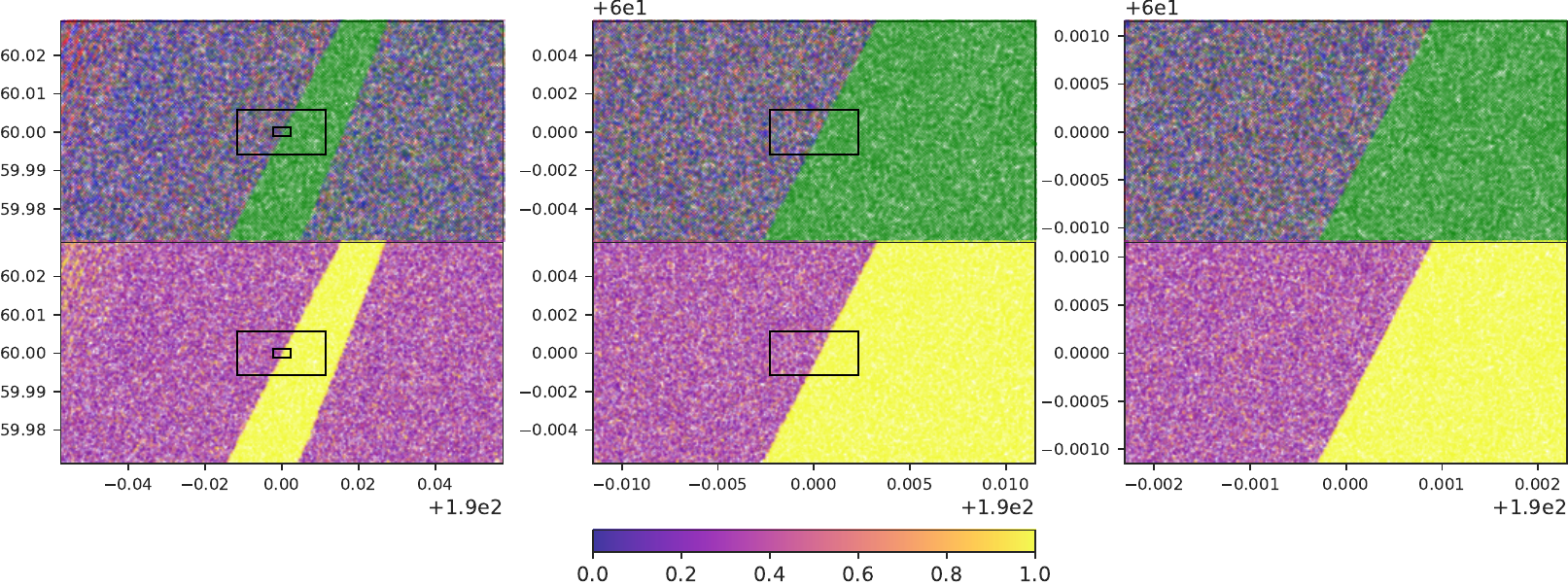}
		\caption{From left to right: initial condition space of the alternative zoom-ins for the zoom levels 5, 6 and 7, in inclination $\iota$ ($y$-axis) and true longitude $\lambda$ ($x$-axis). As in Figure~\ref{fig:bigfig}, the dots are colour-coded by the identity of the escaping particles. Red: $12.5 \,\rm M_\odot$. Blue: $15 \,\rm M_\odot$. Green: $17.5 \,\rm M_\odot$. Bottom panels: fraction of same-colour neighbouring particles $f_\mathrm{col}$, out of $k=9$ nearest neighbours. The black boxes show the boundary box of the next zoomed regions.}
		\label{fig:extrazooms}
	\end{figure*}

	\REV{
		Figure~\ref{fig:extrazooms} displays the initial phase space for alternative zoom-ins at levels 5, 6, and 7. This zoom-in targeted a region at the border between regular and chaotic regions. In the last two zoom-ins, the regular region occupies half of the box, while the chaotic region shows no resolved regular substructures. The measured fractal dimensions indicate the absence of complexity of the regular regions, with values of $D_2=1.88\pm0.00937$, $1.97\pm0.0037$, and $1.97\pm0.00384$ for the three zoom levels, respectively. In the last two zoom levels, $D_2 \simeq D = 2$, reflecting the lack of resolved regular substructures within the chaotic region.
		
		We speculate that with arbitrarily high resolution and numerical accuracy, it would be possible to resolve the regular regions within the chaotic region. Fine regular stripes can be observed on the left side of zoom level 5 (also visible in the outer 4th zoom in Figure~\ref{fig:zoomphase_ejected}), which progressively become finer until disappearing into the chaotic region on the right. These stripes likely correspond to commensurabilities between the return time of the single and the period of the leftover binary. As explained in Section~\ref{sec:fractals}, the regular regions are bordered by extremely long excursions, which can easily accumulate numerical errors due to the extended integration time. These numerical errors can then dephase the coherence of regular regions in phase space (as discussed in Section~\ref{sec:numchaos}), creating the appearance of a chaotic region where regular substructures should instead be present.
		
		These zoom-ins were discarded due to the extensive computational time required to resolve them, since they targeted the border of a regular region. As a result, the 5th zoom-in in Figure~\ref{fig:zoomphase_ejected} was off-centred. Nevertheless, they illustrate the sensitivity of the fractal dimension measurement to the specific region of the initial phase space.
	}
	
\end{appendix}

\end{document}